# Vacuum fluctuations, the size of extra spatial dimensions and microscopic black holes at CERN

Dragan Slavkov Hajdukovic[1]
PH Division CERN
CH-1211 Geneva 23
dragan.hajdukovic@cern.ch
[1]On leave from Cetinje, Montenegro

**Abstract**
We have suggested, that the size of extra spatial dimensions (if they exist) should be related to the quantum vacuum fluctuations; an extra dimension must be sufficiently large to allow appearance of virtual quark-antiquark pairs, which are an inherent part of the physical vacuum in quantum chromodynamics. We argue that the conjecture of extra dimensions with the universal size equal to the reduced Compton wavelength of a pion is a serious alternative to the postulated universal mass in the theory of large extra dimensions (LED). Our conjecture leads to the conclusion that the production of artificial mini black holes at LHC at CERN is unlikely. It is shown that the recent lower limit on the mass of a microscopic black hole (established by CMS collaboration at CERN) may be interpreted as the upper limit (equal to six) for the number of extra dimensions. Additionally, we challenge the current wisdom that mini black holes disintegrate through the Hawking radiation. We point that it might be wrong, if there is a hypothetical repulsion between matter and antimatter; resulting in the creation of particle-antiparticle pairs from the quantum vacuum and matter-antimatter asymmetry in the spectrum of radiation.

## 1. Introduction

Recently, in the framework of the Brane World Scenario, a model with large extra dimensions (LED) was proposed [1-4]. It suggests the possibility of the strong gravitational effects in TeV region; consequently it would be possible to produce mini black holes (MBH) in the Large Hadron Collider (LHC) at CERN.

While the contemporary physics is formulated in three spatial dimensions plus one time dimension, speculations about the existence of extra spatial dimensions, persists already two centuries. It was understood in the nineteenth century, that the additional spatial dimensions would violate the firmly established inverse square law for gravitational and electromagnetic forces. In the 1920s, Kaluza and Klein introduced the idea of the finite (or "compactifed") spatial dimensions as a possibility to avoid any conflicts with observational data. Presently, the extra spatial dimensions are advocated by string theories. In fact, extra six spatial dimensions are required for the most economical and symmetric (i.e. supersymmetric) formulation of a string theory.

Let us point how the LED (as a Brane World Model) 'profits' from the extra spatial dimensions. The key point is that the particles of the Standard Model are assumed to be confined to a three dimensional volume, called a brane, whereas gravity can propagate everywhere, on the brane and in the extra-dimensional volume called the bulk. A new fundamental mass-scale $M_*$ might be introduced and related to the ordinary Planck mass through

$$M_*^{2+n} = \left(\frac{\hbar}{c}\right)^n \frac{M_P^2}{R^n}; \quad M_P \equiv \sqrt{\frac{\hbar c}{G}} \qquad (1)$$

where $n$ is the number of the extra spatial dimensions and $R$ their size (more precisely, the additional dimensions are compactified on radii $R$). Let us observe that in the relation (1), $M_*$ is assumed to have the same value for all $n \geq 1$, and consequently $R$ decreases with $n$.

In the framework of LED, the preferred choice, motivated by solution of the hierarchy problem [1-4], is to identify $M_*$ with the electroweak scale, i.e. $M_* \sim M_{EW} \sim 1 TeV/c^2$. Such a choice implies that mini black holes might be created at LHC. There is nearly a consensus among physicists that a mini black hole would disintegrate through Hawking radiation.

In the present paper we question both conjectures; the universality of $M_*$ in Section 2, and the decay through Hawking radiation in Section 4. In Section 3 we comment limits on the minimum black hole mass, recently established by CMS Collaboration at CERN.

## 2. The size of extra dimensions

Unfortunately, the size of the compact extra dimensions is not fixed in a string theory. It may vary from the size of the Planck length $L_P$ to the size of one nanometre as in the most optimistic scenario with Large Extra Dimensions.

The size of extra spatial dimensions should play a fundamental role in physics and cosmology. Hence, so huge uncertainty (nearly 30 orders of magnitude) in the size of extra dimensions is a major problem. We need plausible physical arguments to limit this uncertainty.

In the present paper, we put forward the idea that there is a universal size ($R_{ed}$) of finite extra dimensions, related to the quantum vacuum fluctuations. More precisely, the conjecture is that an extra dimension must be sufficiently large to allow the appearance of virtual quark-antiquark pairs, which are an inherent part of the physical vacuum in quantum chromodynamics.

To our best knowledge this is the first time that the size of extra dimensions is related to vacuum fluctuations, i.e. demand that extra dimensions are not "sterile" but allow the existence of major fluctuations such as the quark-antiquark pairs.

The size of a virtual particle-antiparticle pair (with inertial mass $2m$) may be approximated with the reduced Compton wavelength $\lambdabar_m = \hbar/mc$. In principle, if $\lambdabar_m > R_{ed}$, virtual pairs with mass $m$ should be prevented in additional dimensions. For instance if an extra dimension has a size $R_{ed} = L_P \approx 10^{-35} m$, it should be "hostile" for virtual electron-positron pairs ($\lambdabar_e = 3.86 \times 10^{-13} m$) or for virtual quark-antiquark pairs. Hence, the assumption that virtual quark-antiquark pairs (for simplicity we consider them as virtual pions) must be allowed, leads to conclusion that the reduced Compton wavelength $\lambdabar_\pi$ of a pion, should be considered as a lower bound for the size of extra dimensions.

We are in a field in which our knowledge is so incomplete that we strongly depend on imagination and guessing. So, let us point out that pions are involved (more than any other particle) in a number of numerical "coincidences" [5-7]. For instance, in the framework of the standard cosmology with dark energy in the form of cosmological constant, the mass of a pion may be related [7] to the fundamental physical constants and cosmological parameters

$$m_\pi^3 = 4\pi \frac{\hbar^2}{cG} H \sqrt{\Omega_\Lambda} \qquad (2)$$

where $H$ and $\Omega_\Lambda$ respectively denote Hubble parameter and dark energy density of the Universe. It must be noted that while $H$ and $\Omega_\Lambda$ change with time, the product $H\sqrt{\Omega_\Lambda}$ is a constant permanently equal to $H_0 \sqrt{\Omega_{\Lambda 0}}$ where index zero, as usually, denotes the present day value. Relation (2) and other numerical coincidences involving pions may be considered as a hint that pions (as the simplest quark-antiquark pairs) somehow dominate the quantum vacuum. Hence, we may guess that the size $R_{ed}$ of an extra dimension may be approximated with the Compton wavelength of a pion.

$$R_{ed} \approx \lambdabar_\pi \tag{3}$$

Now let's assume the existence of $n$ extra spatial dimensions with the size $R_{ed}$ not necessarily fixed by relation (3). The Newton law of gravitation and its analogue in the case of $n$ extra dimensions are

$$F = G \frac{Mm}{r^2} \tag{4}$$

$$F_n = G_n \frac{Mm}{r^{2+n}} \tag{5}$$

Equalling these forces at distance $r = R_{ed}$, gives relation

$$G_n = G R_{ed}^n \tag{6}$$

The familiar Planck mass $M_P^2 = \hbar c / G$ may be easily generalized to the case with $n$ extra dimensions

$$M_{Pn}^{2+n} = \frac{\hbar c}{G_n} \left( \frac{\hbar}{c} \right)^n \tag{7}$$

or using relation (6)

$$M_{Pn}^{2+n} = M_P^2 \left( \frac{\hbar}{c R_{ed}} \right)^n \tag{8}$$

Under the conjecture (3), equations (6) and (8) may be respectively written as:

$$G_n = G \lambdabar_\pi^n \tag{9}$$

$$M_{Pn}^{2+n} = M_P^2 m_\pi^n \tag{10}$$

It is evident that relation (1) and our relation (10) represent two mutually excluding alternatives (i.e. two particular cases of relation (8)); the first one (LED) assuming a unique mass scale $M_{Pn} \equiv M_*$ independent of the number of extra dimensions, and the second one assuming a universal size of extra dimensions. Hence, this opens the question, which conjecture is more plausible; to fix $M_*$ or $R_{ed}$.

For a hypothetical mini black hole with mass $M$, there is a simple relation

$$R_{Sn}^{n+1} = R_S \lambdabar_\pi^n \tag{11}$$

between the familiar Schwarzschild radius $R_S = 2GM/c^2$ and the Schwarzschild radius $R_{Sn}$ corresponding to $n$ extra dimensions. The result (11) is a trivial consequence of the higher dimensional Schwarzschild-metric [8] and conjecture (3).

In principle, the smallest black hole in a space with $n$ extra dimensions should have the corresponding Planck mass $M_{Pn}$ determined by relation (10). The corresponding numerical values presented in Table 1 suggest that the creation of the artificial mini black holes at the Large Hadron Collider at CERN would not be possible. (Let's remember that the maximum centre-of-mass energy that can be achieved at LHC at CERN is 14TeV, which is insufficient if the numbers in Table 1 are a good approximation). This prediction is contrary to expectations raised by the LED phenomenology.

Table 1: Numerical values corresponding to equation (10)

| n | $M_{Pn}$ [kg] | $M_{Pn}$ [$TeV/c^2$] |
|---|---|---|
| 0 | $2.176 \times 10^{-8}$ | $1.2 \times 10^{16}$ |
| 1 | $4.9 \times 10^{-15}$ | $2.7 \times 10^9$ |
| 2 | $2.3 \times 10^{-18}$ | $4 \times 10^4$ |
| 5 | $1.2 \times 10^{-22}$ | 65 |
| 6 | $2.4 \times 10^{-23}$ | 13 |
| 7 | $6.7 \times 10^{-24}$ | 3.6 |
| 8 | $2.4 \times 10^{-24}$ | 1.3 |

Our calculations indicate only the order of the magnitude. So, as Table 1 shows, the creation of mini black holes in six (or more) extra spatial dimensions can't be completely excluded, but in less than six dimensions, the LHC energy is insufficient.

The relation (10) tells us that for very large $n$, $M_{Pn} \approx m_\pi$, while for $n = 0$, $M_{P0} = M_P$.

Let us imagine that the early Universe had a very large number of extra dimensions and that this number has been reduced to the present, unknown value $n_0$ (or alternatively, the number of extra dimensions has grown from zero to $n_0$). Hence the idea is that decrease (or increase) of the number of extra dimensions stops at a certain value $M_{Pn_0}$. Now, let us assume (with the same motivation as in LED) that $TeV/c^2$ is the order of magnitude of $M_{Pn_0}$. If so, according to Table 1, the corresponding value of $n_0$ should be six or seven, what is at least an amusing coincidence with the number of extra dimensions requested by superstring theory.

At the end of this section, let us note that if conjecture (3) is correct, leptons should be suppressed from extra dimensions (with exception of tau, a single lepton more massive than pion).

## 3. The first results at CERN

In December 2010, The Compact Muon Solenoid (CMS) Collaboration at CERN's Large Hadron Collider (LHC) has announced [9] the results of the first dedicated search for microscopic black holes. The search is based on data accumulated by the CMS detector during 2010 from proton-proton (pp) collisions at the center-of-mass energy of 7 TeV. The main result is: "The lower limits on the black hole mass at 95% CL range from 3.5 to 4.5 TeV for values of the Planck scale up to 3.5 TeV in the model with large extra dimensions in space."

Compared with our Table 1, this experimental result indicates that $n = 6$ is the upper limit on the number of extra spatial dimensions.

## 4. Non-thermal disintegration of mini black holes

If mini black holes exist, it is plausible to assume their decay through the Hawking radiation. However we should not neglect the other possibilities and here we point out an interesting alternative.

Let us start with the question: how well do we know the matter-antimatter interactions? It may be surprising, but the available experimental evidence does not allow excluding the hypothesis of the gravitational repulsion between particles and antiparticles (or alternatively, a short range repulsion of non-gravitational origin).

Inside the horizon of a black hole, this hypothetical force might create particle-antiparticle pairs from the quantum vacuum through the gravitational version [10] of the Schwinger mechanism in quantum electrodynamics. While the created particles must stay confined inside the horizon, the antiparticles (because of the assumed repulsion) should be violently ejected. Hence, a black hole made from matter might radiate antimatter and vice versa.

This mechanism of decay should dominate the Hawking radiation, resulting in a shorter life-time of mini black holes. However, the main observable signature of a repulsive force would be presence of more antimatter than matter in the products of disintegration. This should stay valid not only for the semi-classical black holes considered here but for the hypothetical quantum black holes and string balls as well.

If mini black holes decay through the gravitational Schwinger mechanism, intuitively, we expect that evaporation would stop before the complete disintegration; hence a black hole remnant might exist. It would be interesting to find if the hypothetical gravitational repulsion between matter and antimatter is possible in the framework of noncommutative black holes [11], a pioneering approach to quantum gravity predicting a stable black hole remnant.

## *5. Conclusion*

We have proposed a universal size of hypothetical extra dimensions as an alternative to Large Extra Dimensions theory, based on the assumption of a unique mass scale. Our conjecture is motivated by the physical demand that virtual quark-antiquark pairs (as inherent part of the vacuum fluctuations) are not suppressed from extra dimensions and hence the size of extra dimensions can't be smaller than the reduced Compton wavelength of a pion.

Combination of our conjecture and the recent experimental results at CERN, indicate that $n = 6$ is the upper limit on the number of extra spatial dimensions.

**References**

[1] Arkani-Hamed N., Dimopoulos S. and Dvali G.R, *Phys. Lett. B* **429** (1998) 263.
Arkani-Hamed N., Dimopoulos S. and Dvali G.R., *Phys. Rev. D* **59** (1999) 086004.
[2] Antoniadis I., Arkani-Hamed N., Dimopoulos S. and Dvali G.R., *Phys. Lett. B* **4**36 (1998) 257
[3] Landsberg G., *J. Phys. G* **32** (2006) R337
[4] Kanti P., *Lecture Notes in Physics*, Springer Berlin, Volume **769** (2009)
[5] P.A.M. Dirac, *Nature,* **139** (1937) 323
P.A.M. Dirac, *Proc. Roy. Soc., A* **165** (1938) 199
[6] Weinberg S., *Gravitation and Cosmology* (John Wiley and Sons, New York) 1972, p620
[7] Hajdukovic D.S., *Astrophys Space Sci*. **326** (2010) 3
[8] Myers R.C. and Perry M.J., *Annals Phys.* **172** (1986) 304
[9] CMS Collaboration, arXiv:1012.3375v1 [hep-ex] (2010)
[10] Hajdukovic D.S.,arXiv:0710.4316v4 [gr-qc] (2007)
[11] Nicolini P., *International Journal of Modern Physics A* **24**(2009) 1229–1308